\documentclass[sigconf]{acmart}
\usepackage{amsmath,amsfonts}
\usepackage{algorithmic}
\usepackage{graphicx}
\usepackage{textcomp}
\usepackage{xcolor}
\usepackage{booktabs} 
\usepackage{multirow}
\usepackage{placeins}

\AtBeginDocument{%
  \providecommand\BibTeX{{%
    \normalfont B\kern-0.5em{\scshape i\kern-0.25em b}\kern-0.8em\TeX}}}

\begin{document}

\title{``Are you home alone?'' ``Yes''\\
	Disclosing Security and Privacy Vulnerabilities in Alexa Skills}



\author{Dan Su}
\affiliation{%
  \institution{Beijing Jiaotong University}
}
\email{sudan1@bjtu.edu.cn}

\author{Jiqiang Liu}
\affiliation{%
  \institution{Beijing Jiaotong University}
}
\email{jqliu@bjtu.edu.cn}

\author{Sencun Zhu}
\affiliation{%
  \institution{The Pennsylvania State University}
}
\email{sxz16@psu.edu}

\author{Xiaoyang Wang}
\affiliation{%
  \institution{Beijing Jiaotong University}
}
\email{shawnwang@bjtu.edu.cn}

\author{Wei Wang}
\affiliation{%
  \institution{Beijing Jiaotong University}
}
\email{wangwei1@bjtu.edu.cn}

\begin{abstract}
  	The home voice assistants such as Amazon Alexa have become increasingly popular due to many interesting voice-activated services provided through special applications called skills. These skills, though useful, have also introduced new security and privacy challenges. Prior work has verified that Alexa is vulnerable to multiple types of voice attacks, but the security and privacy risk of using skills has not been 
  fully investigated. In this work, we study an adversary model that covers three severe privacy-related vulnerabilities, namely, \textit{over-privileged resource access}, \textit{hidden code-manipulation} and \textit{hidden content-manipulation}. By exploiting these vulnerabilities, malicious skills can not only bypass the security tests in the vetting process, but also surreptitiously change their original functions in an attempt to steal users' personal information. What makes the situation even worse is that the attacks can be extended from virtual networks to the physical world. 
	We systematically study the security issues from the feasibility and implementation of the attacks to the design of countermeasures. 
	We also made a comprehensive survey study of 33,744 skills in Alex Skills Store.
\end{abstract}

\keywords{Internet of Things (IoT) security, security and privacy, voice assistant security}

\maketitle

 \section{Introduction} 

Home voice assistants, e.g., Amazon Alexa and Google Assistant, have gained tremendous popularity in recent years. In December 2019, smart speaker households in the U.S. own 2.6 smart speakers on average with total smart speakers in use rising to 157 millions \cite{smart-speaker-ownership}. Amazon Echo, which is one of the Alexa-enabled devices, maintains a 69.7\% market share among smart speakers \cite{market-share}. Alexa offers simple access to perform everyday tasks and responds naturally to voice commands such as ``Alexa, what's the weather? ''. These functions are implemented by multiple applications called \textit{skills} by Amazon. In addition to performing built-in functions, Alexa encourages the third-party developers to build new skills as a way of expanding its service areas. This open ecosystem greatly enhances Alexa's usability and improves users' experience. Alexa skill count has surpassed 100,000 worldwide \cite{Number-of-skills-2019}, a massive increase from only 1,000 skills in June 2016, 
and the number is still growing. Alexa has become one of the most popular smart home products, but in the meantime concerns arise: \textit{Is it safe to let Alexa know everything about us?}

Users login to Alexa with Amazon accounts, which contain users' personal information such as their names, phone numbers, shipping addresses and emails. The information can be accessed by skills. As mentioned above, Alexa has two types of skills: official built-in skills and  third-party skills. Users may trust the official skills, but whether the third-party skills can secure their information remains a question because the third-party skills can be developed by adversaries. Similar to the Android ecosystem where malicious apps frequently steal users' private information, malicious skills may also steal users' personal information. In addition, most Alexa devices are placed at home and can be accessed by anyone in the family, including kids. The latest news has reported that a woman suffered from violence threat because Alexa told her to ``kill herself to save the environment'' \cite{Alexa-news-kill}. In this work, we are motivated to explore the reasons that Alexa occasionally behaves weird by in-depth analysis of Alexa's third-party skills. We establish an adversary model that contains three severe vulnerabilities:

\textbf{Over-privileged resource access:} In order to provide relevant and satisfactory services, some skills require users' personal information, e.g., a weather skill requires users' location. There are two ways to realize the purpose: one is to request relevant permissions when users enable the skills, and the other is to directly ask users during real-time conversations when users are using the skills. The acquired users' information can be further sent to remote servers and then collected by developers. We observed that some skills are able to work correctly and identically no matter the requested permissions are granted or not. In other words, users' personal information is not the necessity of the skill's function at all. The designs of these skills have violated the principle of least privilege. Although Amazon's rigorous certification process has paid much attention to permissions, over-privileged skills still exist. In addition, we implemented a novel attack to bypass the vetting process provided by Amazon market and successfully make the data-stealing behavior ``legitimate'' by providing a reasonable excuse in the skill description. It is extremely risky if all the personal information can be easily accessed by attackers. 

\textbf{Hidden code-manipulation:} The vulnerability is due to Alexa skills' development mechanism. A skill's development can be divided into two modules: frontend and backend. Specifically, the frontend provides the skill's interaction model that defines the requests the skill can handle and the utterances users make to invoke these requests. The backend provides a cloud-based service that handles the requests. The frontend and the backend are closely linked to achieve the normal operation of the skill. However, according to our tests and previous work \cite{wangxiaofeng}, in general Amazon does not conduct a second round review in a long time after the skill is released to the skill market as long as the frontend remains the same. Taking advantage of this, attackers can easily manipulate the skill's functions by modifying only the backend code. We have successfully changed the skill's real-time conversation from a common question ``Do you want to hear a joke?'' to a sensitive question ``Are you home alone?'', while maintaining the frontend utterances. A ``Yes'' answer from the users along with their home address information will enable attacks to transfer from the Internet to the physical world.

\textbf{Hidden content-manipulation:} News skills are skills that provide daily headlines. A news skill requires a feed as the news source. Due to the consideration of diversity, Amazon does not make stringent requirements on the feed: either official news or personal blog website is accepted. Instead of publishing a news skill directly linked to a malicious website and thus getting rejected during the vetting process provided by Amazon market, attackers can first link the skill to their own websites that show normal content. Then they remotely conduct the attack by manipulating the content of the website. The biggest difference between code-manipulation and content manipulation is that the latter does not modify the code, which makes the attack highly stealthy and can only be discovered when the skill is activated. What's more, the attack is \textbf{not} specific to News skills. All custom skills can implement the function of parsing the feed and read messages from websites, which makes them potential threats to users. Attackers are able to add improper content including but not limited to pornography, violence, advertisement and fake news, which in consequence seriously violates Amazon's regulations and potentially causes terrible social effects. To the best of out knowledge, this is the first work to present the vulnerability on Alexa skills.

Analyzing the security of Alexa skills faces some great challenges. First, code analysis is infeasible. A skill is a cloud-based service that is hosted on the Amazon's or developer's personal server, so we are unable to access its source code directly to determine whether its behaviors are legitimate or not. Second, network traffic analysis is difficult. Users and the skill servers are isolated by Amazon cloud. On the one hand, we are unable to monitor the network on skill server's side, so we cannot track down what exactly the developers do with our personal information. On the other hand, on user's side, the network flows are encrypted and the mobile Alexa app is designed to prevent man-in-the-middle https proxy (e.g., \textit{fiddler}) based traffic analysis. Consequently, it is nearly not possible to analyze the plaintext content of network flows with existing tools. Third, it is difficult to collect a grand scale of skills' information for research. Previous research on Alexa skills was mainly on voice attacks. For example, Zhang \textit{et al.} \cite{wangxiaofeng} and Kumar \textit{et al.}\cite{DBLP:conf/uss/KumarPMHMBB18} presented skill squatting attacks. To the best of our knowledge, this is the first systematic work focusing on the disclosure of privacy and security vulnerabilities in Alexa skills.

In this work, we make the following contributions: 
\begin{itemize} \setlength{\itemsep}{-0.5ex}
	\item  We present an adversary model that contains three severe vulnerabilities in Alexa skills: over-privileged resource access, hidden code-manipulation and hidden content-manipul-ation. We successfully exploited these vulnerabilities with attacks and bypassed Amazon's vetting process, including function test, voice interface test, policy test and security test. We provide a real scenario to illuminate that the attack can extend from the virtual network to the real world by exposing users to physical threats.
	
	\item We conduct an in-depth study of Alexa's ecosystem, skill's development and Amazon's vetting process to understand the root causes of the attacks. We also provide feasible countermeasures to the vetting process.
	
	\item We conduct a thorough survey study of 33,744 skills in the market to collect their requested permissions, invocation names and developers. We provide comprehensive statistics and analysis that none of previous work has presented. We found 57 over-privileged and 11 potentially over-privileged skills in the market.
\end{itemize}

\section{Background}
\label{background}

\subsection{Alexa's ecosystem} 

Amazon Alexa is a virtual assistant with the capability of interacting with users through voice instead of traditional computer interfaces like keyboard or mouse. It is capable of playing audiobooks, ordering food, controlling smart devices or providing real-time information on weather, traffic and sports. Users can extend Alexa capabilities by enabling different skills. 

\begin{figure}[t]
	\centering
	\setlength{\abovecaptionskip}{0cm}
		\setlength{\belowcaptionskip}{-0.15cm}
	
	\includegraphics[width=0.45\textwidth]{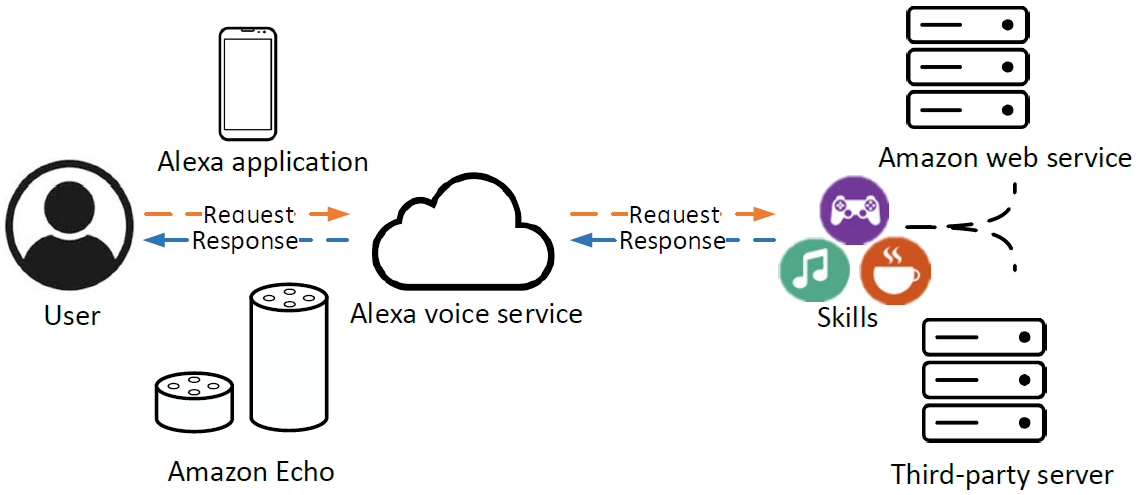}
	\caption{Alexa ecosystem}\label{ecosystem}
\end{figure}

The Alexa ecosystem consists of four components: user, device, Alexa voice service (AVS), and skill servers. The device's job is to take voice commands as input and to output skills' responses. Users give a voice command to a device, e.g., Amazon Echo or a mobile phone equipped with the Alexa app. The device passes the voice stream to AVS where the stream is analyzed. The main function of Alexa is implemented by AVS that resides in the cloud instead of being built into an Amazon Echo. AVS identifies the requested skill with Natural Language Processing and creates a request that includes the requested skill and the intent. Then the request is sent to the skill server and handled by the requested skill's intent handler. The text response gets back to AVS where the text is converted into voice stream by the Text-to-speech system. Finally the response is played by the Alexa-enabled device to the user. Fig.~\ref{ecosystem} shows details of the Alexa ecosystem.

\subsection{Alexa skills' development} 
Alexa provides a set of built-in capabilities, referred to as skills, including but not limited to playing music, making to-do list and answering questions. Third party developers can build new skills with Alexa Skills Kit (ASK), which provides APIs, tools and documentation. Developers can either choose to create a skill in minutes using the skill blueprint templates or design a custom skill by themselves. For example, developers can use the \textit{flash briefing template} to create a news skill. The only information they should provide is a news feed link.

For custom skills, the development includes the frontend and the backend. The frontend can be built with ASK. Intents, utterances, invocation names permissions and endpoint link are necessary in the frontend setting in ASK~\cite{Build-Alexa-skills}. Details are shown in Appendix ~\ref{appendix_elements}.

Besides the frontend setting within ASK, developers should also build the skill's backend. The backend is a cloud-based service that can handle all the intent requests defined in the frontend and implement relevant functions. The backend can be hosted on Amazon Web Service (AWS) or on developers' own servers. The frontend and backend are linked with the element called \textit{Endpoint link} in ASK. 

\vspace{-0.2cm}
\subsection{Certification} 
\label{certification}
When submitting a skill, the skill must pass several tests during the Amazon vetting process before it can be released, including functional test, voice interface test, policy test and security test \cite{certification-requirements}. The details of tests are shown in Appendix ~\ref{appendix_cer}.

Developers can check the vetting status via the ASK console. They are expected to revise the skill according to reviewers' feedback if the skill fails the tests. Otherwise, the skill can be released online on the skill market. According to our observation and previous work \cite{wangxiaofeng}, Amazon does not take the second round of vetting in a long time except in some special situations, e.g., when offensive or pornographic words are detected. There seems to be a blacklist of words monitored by Amazon's Text-to-speech system. However, the responses were different from time to time. 
For a while Alexa made a beep sound instead of reading the bad words, and for a while the beep sound disappeared and the bad words were read normally. The differences were probably caused by Alexa's updates.

\vspace{-0.2cm}
\subsection{Comparison of Alexa skills and Android mobile apps}  
Alexa skills and Android mobile apps have a lot in common. They are both developed to expand the capabilities of the devices. Similar to Android apps, there exist official skills and the third party skills. Similar to Google Play app store, in Amazon skill market users can also access skills' descriptions and enable them (no actual downloading and installation though). In terms of vetting, both skills and apps should pass the vetting process before being released to the market. 

However, there are also significant differences between them, which makes the distribution model of skills different from that of Android apps : an Android app is a compressed Android Package (APK) file that contains the app's configuration file, code, resources, and certificates. When an Android mobile app is being published, all these files are available and transparent for review, even if some of them are obfuscated. However, a skill is a cloud-based service. Users can enable a skill with a simple click instead of downloading it since it does not have an entity shown to users. When reviewing a skill, the reviewer can only access its frontend, which defines the utterances for invoking different intents. The backend source code hosted by the developer is a black box to the reviewer. The vetting is only based on the skill's real time response. More specifically, if an Android app needs an update, a new version APK should be uploaded to the market and get tested. But for a skill, the vetting does not take place unless the skill's frontend changes. In other words, attackers are able to modify the backend code without being noticed by Amazon. They can submit a benign skill for vetting. After the skill is judged as legitimate and published in the market, they may change the backend code to add malicious content or change the skill's functions without editing the frontend. It is thus extremely risky to the Alexa ecosystem if a large number of skills' actions do not match their descriptions. We provide a detailed illustration of the attacks in the next section.

\vspace{-0.2cm}
\section{Adversary Model} 
\begin{figure}[t]
	\centering
		\setlength{\abovecaptionskip}{0.1cm}
		\setlength{\belowcaptionskip}{-0.1cm}

	\includegraphics[width=0.45\textwidth]{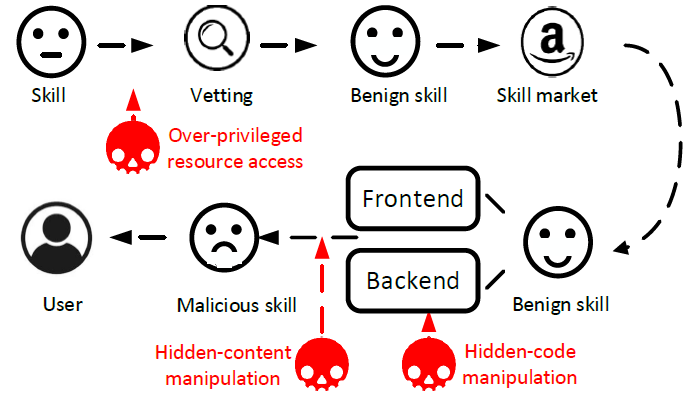}
	\caption{Adversary model}\label{model}
\end{figure}

The attackers do not access to the Alexa cloud and the Alexa-enabled devices. Victims are the Alexa skill users. Three attacks could take place: (1) For the purpose of collecting users' personal information, attackers may request more permissions or ask well-designed sensitive questions in the skill. (2) Utilizing the drawback that Amazon only reviews the frontend of the skill during the vetting while the backend is not accessible, hidden-code manipulation attack may occur by changing the backend code while keeping  the frontend intact. (3) Taking advantage of the lack of periodically reviews, attackers can inject harmful content to the skill for a period of time. The manipulated content will not be noticed until the skill is activated by users. The attackers are considered to publish a large number of malicious skills to access as many users as possible. The adversary model is described in Fig.~\ref{model}.

To demonstrate how easy it is to conduct the above attacks, we develop and submit a non-malicious skill to the market.  
We are aware of the ethical concerns for publishing the skill because some users may accidentally enable it. However, we could not obtain an official ethics review because currently there is no such committee in our university yet. Therefore, we have tried our best to follow the common practices to minimize privacy risks.
Specifically, in order to make our experimental skill most probably for our research use only, we give the skill a non-intuitive name (``susu assistant''), so that general users will not notice and use it. In addition, we disable the skill's functions when no evaluation is in process. What's more, the skill was briefly available in the store and has been removed after the experiments.


\vspace{-0.2cm}
\subsection{Over-privileged resource access} 
To provide relevant services, a skill may require users' specific information, e.g., a weather skill needs users' locations to provide weather reports. Based on our observation, skills can access users' information by either requesting permissions or asking users in the skill.  In the first case, a permission checkbox pops up when users enables the skill. If users agree to grant the permissions, the skill can directly access the users' Amazon account information. In the second case, developers define an Intent with certain kinds of slots to receive users' messages; for example, \textit{The All American Steakhouse Ashburn} is a food skill which can be used to order food or reserve tables. Though it does not request the permission of user's phone number, it asks users in the skill ``What's your phone number?''. Amazon provides multiple predefined slots that can be employed to trigger the Intent, e.g., the slot for phone numbers is \textit{AMAZON.PhoneNumber}.

In the function test of the vetting process, Alexa tests whether a skill can correctly work in both conditions when users have or have not granted the requested permissions. However, Alexa does not have stringent restrictions on what a skill may request permissions for. That is to say, Alexa does not evaluate the quality of the function that needs the permission. Since no previous work has focused on the permissions of skills, and the permission mechanism is similar to that of the mobile platforms, we use the definition of \textit{over-privileged} as ``request more permissions than it requires '' \cite{Permission-basedSecurityModels, AndroidPermissionsDemystified, PScout}.

\begin{figure}[htbp]
\setlength{\abovecaptionskip}{0.1cm}
		\setlength{\belowcaptionskip}{-0.1cm}

	\centering
	\includegraphics[width=.45\textwidth]{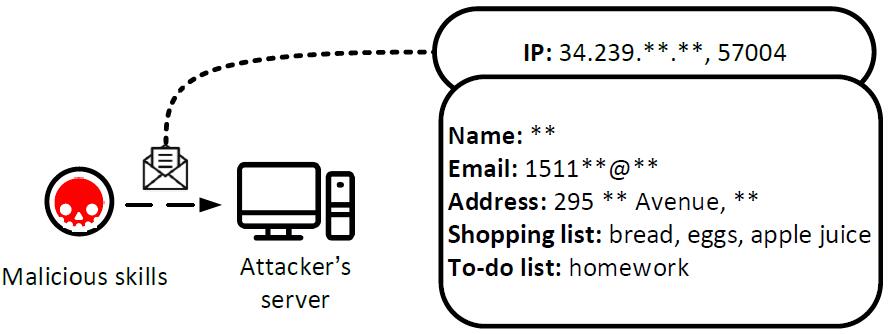}
	\caption{User's information is sent to our server (here the user is an author of this paper). The information is anonymized.}\label{server_data}
\end{figure}

Unfortunately, well-designed skills are still able to make the behavior of over-privilege legitimate. For the purpose of passing the vetting and successfully collecting users' information, attackers can develop a ``useless'' skill which has a reasonable excuse in the skill description. To prove this, we developed a skill called \textit{susu assistant} that requests permissions of almost all the user's information, including full name, phone number, email, address, shopping list and to-do list. The only thing the skill does is to tell the users their own information. Users can ask ``What's my phone number?'' and the skill responds ``Your phone number is \ldots''. In the skill description, it says ``The skill is designed for the elderly or people who always forget their own information''. The skill successfully passed the vetting by Amazon. As a test user of this skill, we enabled the skill and saw our information was transferred to our server, as shown in Fig.~\ref{server_data}. 

\begin{figure}[htbp]

\setlength{\abovecaptionskip}{0.1cm}
		\setlength{\belowcaptionskip}{-0.1cm}

	\centering
	\includegraphics[width=.42\textwidth]{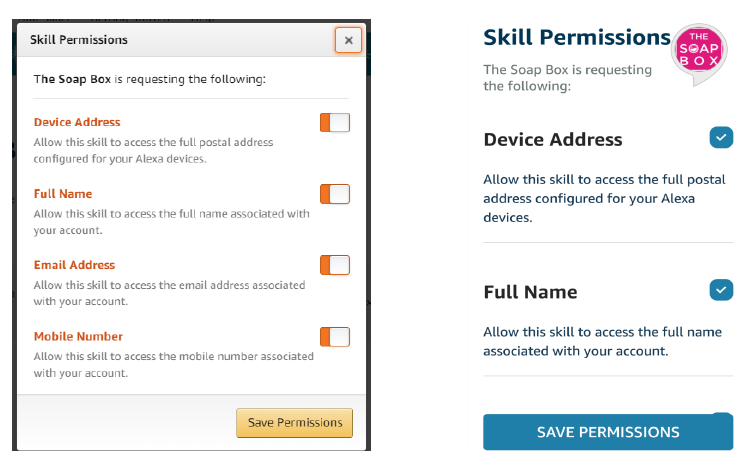}
	\caption{The default status of the permission checkbox is checked when enabling the skill via Alexa website and via Alexa app. In the case that the user enables the skill by saying the skill's name via voice command, the skill's permissions are not granted by default. However, this is not suitable for new users since they do not know the name. }\label{checkbox}
\end{figure}

What makes the situation even worse is that the default status of the permissions' checkbox is checked when users click the "enable" button of the skill, as shown in Fig.~\ref{checkbox}. For ordinary users who lack professional knowledge and privacy awareness, when they enable our skill, they may automatically agree to grant all the permissions without understanding the risk. 
Even if they decide to disable such a malicious-purpose skill after using it for only a few seconds, their personal information has already been stolen by attackers. 

Skills that request permissions are supposed to provide a privacy policy displayed to users on the skill's webpage. Although in security test of the vetting process, Amazon has the privacy requirements that developers should comply with the privacy policy, we do not find any specific measures to ensure developers follow the agreement. A number of websites can help create a seemingly formal privacy policy in minutes, e.g., \textit{www.freeprivacypolicy.com}. The content claims that the collected data is used for providing better services and is under good protection by \textit{susu assistant}. However, it is unable to guarantee that. Besides, lots of skills' privacy policies claim that the skills (e.g.,\textit{flash card quiz for citrix ports}) require users' information, but they do not request anything in reality. Both situations illuminate that the privacy agreement does not provide practical guidance on verifying the skill's legitimacy.

\vspace{-0.2cm}
\subsection{Hidden code-manipulation}  
As discussed in Section~\ref{background}, after the skill has been published to the market, Amazon will not take the second round vetting in a long time. Developers can surreptitiously manipulate the backend code without changing anything in the frontend, thus evading Amazon's detection. Richard et al \cite{mitev2019alexa} mentioned the attack in their work, however, details of root reasons and implementations have not been discussed. Here, we provide detailed attack scenarios to explain how the attack can extend from virtual networks to the real world.

To simulate the attack scenario, we again take \textit{susu assistant} as an example. Firstly, we defined two new intents in the skill's frontend, \textit{YesIntent} and \textit{NoIntent}. The utterances associated with the two intents are [``Yes'', ``OK'', ``Sure''] and [``No''], respectively.
During the conversation, the skill asks users whether they would like a joke by saying ``Do you want to hear a joke?'' If the users agree and say any one of ``Yes'', ``OK'' or ``Sure'', the \textit{YesIntent} will be invoked and the skill will tell users a joke. Otherwise, if the users respond with ``No'', the \textit{NoIntent} will be invoked and the skill will perform other tasks. 

\begin{figure}[htbp]
\setlength{\abovecaptionskip}{0.1cm}
		\setlength{\belowcaptionskip}{-0.1cm}

	\centering
	\includegraphics[width=.5\textwidth]{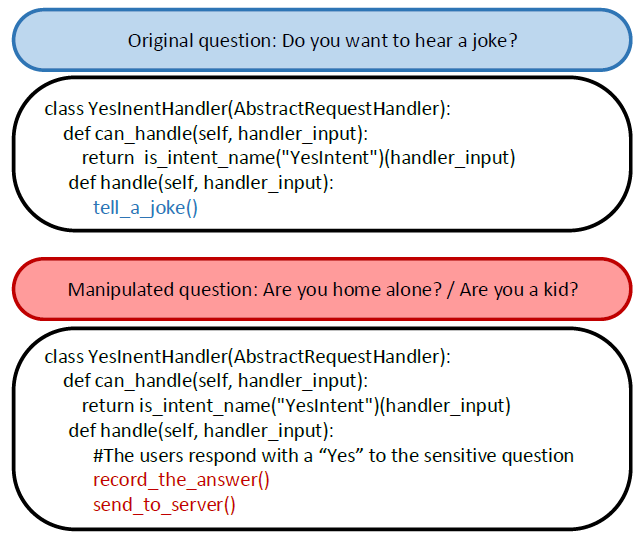}
	\caption{Manipulated intent handler}\label{intent_handler}
\end{figure}

\begin{figure}[htbp]
\setlength{\abovecaptionskip}{0.1cm}
		\setlength{\belowcaptionskip}{-0.1cm}

	\centering
	\includegraphics[width=.45\textwidth]{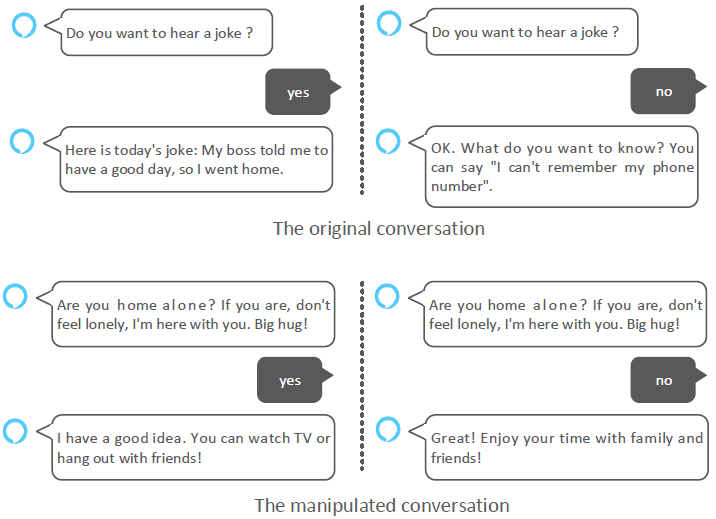}
	\caption{The original and manipulated conversations}\label{conversation}
\end{figure}

Secondly, we submitted the skill. It passed the vetting and became online in the market. Thirdly, we manipulated the backend code while preserving the intents' utterances in the frontend. We changed the context from ``Do you want to hear a joke?'' to several personal questions like ``Are you home alone?'', ``Are you over 18?'' or ``Are you a girl?''. We also changed the handlers of \textit{YesInent} and \textit{NoIntent} in the backend code, as shown in Fig.~\ref{intent_handler}. When the intent is invoked, the skill records ``yes'' or ``no'' according to users' answers and the information will be further transferred to our server. Since we did not change anything in the frontend, the manipulated version of the skill did not go through the vetting again. The original and the manipulated conversation are shown in Fig.~\ref{conversation}. 

The danger of the hidden code-manipulation attack lies in two aspects: (1) similar to \textit{susu assistant}, the skill may have obtained users' other private information via permissions, e.g., address and phone number. The conversation enables attackers to get real-time information about the users. If the users give ``yes'' to the question ``Are you home alone'', they may be in a dangerous situation as their home addresses have been disclosed via the first attack; (2) Even though it is the user's choice whether to give a real answer not, the attacker may still seduce users by designing a warm context and showing care to eliminate their wariness, e.g., ``Are you home alone? If you are, don't feel lonely, I'm here with you. Big hug!''. There is a big chance to increase the probability of receiving real answers. The situation may be even more dangerous if children invoke the skill. Note that while this attack scenario may appear to any user, our testing only lasted for a short time to demonstrate the feasibility and we were the only ``victim'' user.

Another attack scenario is to add messages related to advertisements, election speech, fake news or rumors. The messages can be hard-coded into the backend code. Attackers can simply add an advertisement in the welcome message. It violates the policy that ``the advertisements should not use Alexa's voice or a similar voice''. Election speech or fake news can deliver misleading information to users, interfere with their choices, or even cause social panic.

\vspace{-0.2cm}
\subsection{Hidden content-manipulation} 
Hidden content-manipulation attack is similar to the second scenario of the hidden code-manipulation attack, which surreptitiously manipulates and delivers special content to users, violating Amazon's regulations.  However, it does not manipulate the background code, which is the biggest difference from code-manipulation attack. The hidden content-manipulation is conducted remotely and it is not limited to specific skills. 
Take News skills as an example. To create a News skill that provides daily news, developers are required to provide a news feed. The feed can be either in JSON format or RSS format. Taking RSS as an example, there are multiple free online RSS makers available, e.g., \textit{Feed43.com}. Developers can make their own RSS feed linked to any web page. However, if a skill is directly linked to a malicious website, e.g., a pornographic website, it will be rejected with the reason ``Your skill includes pornographic images or other content that violates our Content Guidelines''. Thus to bypass the vetting process, attackers can submit a normal skill linked to his own webpage with appropriate content, which they are able to manipulate freely at a later time. There is even no cost at all if attackers host the webpage on GitHub. After the skill passes the vetting and becomes live, the attacker can manipulate the content on his webpage without changing RSS link or anything related to the skill. Thus both frontend and backend of the skill remain the same, evading the vetting process. In addition, unless there exist pornographic information or bad words which can be easily detected via a blacklist, fake news and rumors are difficult to be discovered.

To verify the attack and measure what kinds of inappropriate content may trigger the second-round vetting, we developed 6 skills using the flash briefing template to provide daily jokes. At first, the feeds were linked to 6 different webpages with the joke content. The skills passed the vetting on December 22, 2019. Then we immediately manipulated the jokes on the webpage to 6 kinds of contents, including advertisements, voting mobilization, fake news, rude words, pornography contents, and political views. Unexpectedly, none of the skills have been detected and removed within 7 days. The improper contents may cause terrible social effets. For ethical concerns, the names of skills are nonintuitive so that general users will not notice and use them. The skills are only activated by us for research purpose and have been removed immediately after the tests.

\vspace{-0.2cm}
\subsection{Regulation violations} 
The above attacks are able to break the four tests in Amazon's certification process described in Section~\ref{certification}.  Over-privileged resource access intends to collect users' personal information, resulting in breaking policy test and security test. Hidden code-manipulation violates the function test by manipulating the backend code to change the skill's function. It can also add impolite and offensive remarks, which breaks voice interface test. By broadcasting advertisements in the skill with Alexa's voice, it breaks the policy test. It utilizes the Yes/No question to collect users' information, violating the security rules. Hidden content-manipulation changes the skill's function by manipulating the content of the website. For instance, it may start to be a joke skill which belongs to Game or Novelty category. After manipulating the content, it can provide news which should be categorized in the News category even the news may be fake. Alike hidden code-manipulation, it breaks the voice interface test and policy test. The summary of the violations is shown in Table~\ref{Violations}. Amazon's development regulations are designed to establish application standards and protect user privacy, but now they are arbitrarily violated.

Besides requesting permissions and asking directly in the skill to collect users' personal information, attackers can also ask Yes/No questions to get users' real-time information. Requesting permissions is the easiest to implement and the least alerting, but it achieves the highest precision of information. The comparison of the three approaches is shown in Table~\ref{Comparison of methods}.

\begin{table}[htbp]
	\setlength{\abovecaptionskip}{0cm}
		\setlength{\belowcaptionskip}{-0.15cm}

	\caption{Violations}
	\begin{center}
	\scalebox{0.9}{
		\begin{tabular}{ccccc}
			\toprule  
			&Function&Voice interface&Policy&Security\\
			&test&test&test&test\\
			\midrule
			Attack 1&--&--&$\surd$&$\surd$\\
			Attack 2&$\surd$&$\surd$&$\surd$&$\surd$\\
			Attack 3&$\surd$&$\surd$&$\surd$&--\\
			
			\bottomrule 
			\multicolumn{5}{l}{Attack 1: Over-privileged resource access}\\
			\multicolumn{5}{l}{Attack 2: Hidden code-manipulation}\\
			\multicolumn{5}{l}{Attack 3: Hidden content-manipulation}
		\end{tabular}
		}
		\label{Violations}
	\end{center}
\end{table}

\begin{table}[htbp]
	\setlength{\abovecaptionskip}{0cm}
		\setlength{\belowcaptionskip}{-0.2cm}
		
	\caption{Comparison of three approaches for collecting users' information}
	\begin{center}
	\scalebox{0.9}{
		\begin{tabular}{p{0.5pt}p{57pt}ccc}
			\toprule  
			&&Requesting&Asking directly&Yes/No\\
		
			&&permissions&in the skill&questions\\
			
			\midrule
			1&Vulnerability&Attack 1&Attack 1&Attack 2\\
			2&Users' attention&Low&High&Medium\\
		
			3&Amount of&High&Medium&Low\\
			&information&&&\\
			4&Complexity of&Low&High&Medium\\
			&implementation&&&\\
			
			\bottomrule 
			\multicolumn{5}{l}{Attack 1: Over-privileged resource access}\\
			\multicolumn{5}{l}{Attack 2: Hidden code-manipulation}\\
			\multicolumn{5}{l}{*The values ``High'', ``Medium'' and ``Low'' are relative.}\\
			
		\end{tabular}
		}
		\label{Comparison of methods}
	\end{center}
\end{table}

\vspace{-0.2cm}
\section{A Survey Study of Alexa Skills Market}
It is insecure for users to be exposed to these three types of attacks. To the best of our knowledge, however, there is no related work that provides a survey study on the skill market from the perspective of our adversary model. 
We are motivated to figure out whether there exist malicious skills or exploitable skills in the skills market.

\vspace{-0.2cm}
\subsection{Skills in our dataset} 
Each skill has an introduction webpage in Amazon skills market. We crawled 37,350 skills' pages from 20 categories on August 29, 2019, including invocation names, descriptions, utterances, developers, requested permissions, ratings and reviews. The information provides a holistic view of the skills. The details of the dataset are described in Appendix ~\ref{appendix_dataset}. We notice that many skills have the same skill names. In addition, a number of skills belong to two categories, e.g., a sport skill that provides the latest news about sports belongs to both News and Sport categories. To uniquely identify each skill, we utilized the hash value of a skill's name, invocation name, developer and utterances. Finally we have the meta information of 33,744 unique skills after filtering out reduplicated skills.

To efficiently enable and disable the large number of skills for testing, we used AndroidViewClient~\cite{AndroidViewClient} and adb~\cite{Adb} commands to trigger the widgets in the layout of Alexa apps on an Android smart phone. It can automatically search the skills by name and click the enable/disable button, making the testing process automatic.  

\vspace{-0.2cm}
\subsection{Permission evaluation}  
\label{over-per}
In the function test of the vetting process, Amazon tests whether a skill can appropriately respond in both conditions when users have granted or denied the requested permissions. On the one hand, if the users have granted the requested permissions, all intents should work properly. On the other hand, if the users have denied the permissions, the intents that \textit{do not} need the permissions should work properly, and the intents that \textit{do} need the permissions should remind users to grant permissions. In other words, if all intents can work properly and meanwhile all responses remain the same no matter users grant or deny the permissions, an over-priviledged resource access has occurred. The definition of over-privileged here is the same as that used in privacy leak analysis of mobile apps~\cite{Permission-basedSecurityModels, AndroidPermissionsDemystified, PScout}.
The skills that violate the principle of least privilege are considered over-privileged.
There are ten permissions defined in Alexa. In the following, we only focus on four sensitive permissions that are most closely related to users' personal information: full name, address, phone number and email.

Firstly, we calculated the number of requested permissions by each skill. Secondly, for the skills that request at least one sensitive permission, we tested each of its functions by saying the utterances provided from the skill's webpage as well as those extracted from the internal voice-based help messages. For example, if a skill requests permissions of name and address, we will invoke the skill \textit{at least} three times: for the first time, we do not grant it any permissions; for the second time, we only grant it the name permission; for the last time, we grant it both name and address permissions. We triggered all commands the skill provides and compared the responses of each test.  We also tested each skill many times in September, October and December of 2019 to avoid contingency and ensure the consistency of the conclusion.
If all the intents can work in both conditions and the skill responses the same in each differential test, we determine that it is an over-privileged skill. From Table~\ref{permission}, we have the following observations: 

\begin{table}[htbp]
	\setlength{\abovecaptionskip}{0.1cm}
		\setlength{\belowcaptionskip}{-0.2cm}

	\caption{Number of skills requesting sensitive permissions}
	\begin{center}
	\scalebox{0.9}{
		\setlength{\tabcolsep}{1mm}{
			\begin{tabular}{ccccc}
				\toprule  
				Number of&Total&Available&Over&Potentially\\
				requested&&&-privileged&over\\
				sensitive&skills&skills&skills&-privileged\\
				permissions&&&&skills\\
				\midrule
				4&16&13&1&4\\
				3&23&22&1&0\\
				2&41&37&6&6\\
				1&\multirow{2}*{4}&\multirow{2}*{4}&\multirow{2}*{0}&\multirow{2}*{0}\\
				(Phone number)&&&&\\
				1&\multirow{2}*{3}&\multirow{2}*{3}&\multirow{2}*{0}&\multirow{2}*{0}\\
				(Full name)&&&&\\
				1&\multirow{2}*{32}&\multirow{2}*{27}&\multirow{2}*{2}&\multirow{2}*{1}\\
				(Email)&&&&\\
				1&\multirow{2}*{219}&\multirow{2}*{171}&\multirow{2}*{47}&\multirow{2}*{0}\\
				(Address)&&&&\\
				Total&338&277&57&11\\
				\bottomrule 
				
			\end{tabular}
			}
		}
		\label{permission}
	\end{center}
\end{table}

(1) 338 skills have requested sensitive permissions. However, some of them have functional problems that are not accessible or have been removed from the market, resulting in 277 available skills in the end.

(2) 57 skills are found over-privileged. For example, \textit{Thingee Tech Talk} is a skill that brings the latest advances in digital technology. It requests permissions of full name, phone number and email. However, it works and responses normally even without these permissions; 
hence, it is over-privileged. Similarly, \textit{Diginomica Podcast} plays podcasts from diginomica.com, but it does not use the information of address and full name it requested. What's more, \textit{Western Drought Tracker} provides information about the rainfall and drought conditions in California and Nevada, no matter whether users grant it the address permission or not.  

\begin{figure}[htbp]
\setlength{\abovecaptionskip}{0.1cm}
		\setlength{\abovecaptionskip}{0.1cm}
		\setlength{\belowcaptionskip}{-0.1cm}

	\centering
	\includegraphics[width=0.45\textwidth]{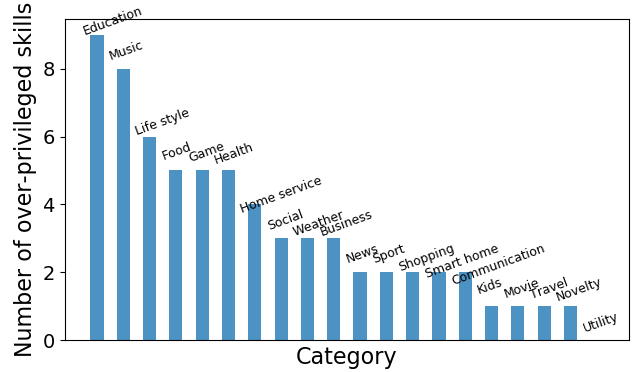}
	\caption{Number of over-privileged skills}\label{over-privileged}
\end{figure}

The distribution of the over-privileged skills is shown in Fig.~\ref{over-privileged}, which gives us an intuitive understanding of which category is more likely to request personal information. Although 57 skills are over-privileged, here 65 skills are shown. The reason is that 8 skills belong to two categories, e.g., \textit{Western States}, a skill that keeps runners up to date on the Western States Endurance Run, belongs to both the Sport and News categories. The Education category has 9 over-privileged skills, making it the category with the highest rate of violations. 

\begin{figure}[htbp]
\setlength{\abovecaptionskip}{0.1cm}
		\setlength{\belowcaptionskip}{-0.1cm}

	\centering
	\includegraphics[width=0.35\textwidth]{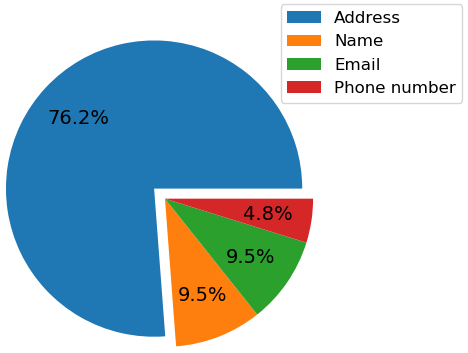}
	\caption{Overused permissions}\label{over-privilege-per}
\end{figure}

The distribution of permissions that are overused is shown in Fig.~\ref{over-privilege-per}. Address is the most overused information, followed by name, email and phone number. Since users log into Alexa with their Amazon accounts, it is a remarkable fact that the address obtained by permissions is the shipping address (normally home address) set in user's Amazon account. The risk is that the home address information may be misused by attackers with evil-purpose (e.g., making more realistic scams targeting at victim users, or even causing safety concerns). 

(3) 11 skills are potentially over-privileged. The skills are considered to be potentially over-privileged because the permission checking is conditionally triggered. For example, there are a number of coupon skills in the market. If coupons are available, the skill may send the coupon information to users by emails or short text messages. Just before sending the coupon, the skill checks whether it has the phone number permission. If it does not have the permission, the skill will output an utterance to ask for the permissions. However, if there is no coupon, the skill only returns ``We don't have coupons at the moment'' and does not give the permission reminder. Therefore, if a coupon skill always has no coupon, its responses are always the same no matter they have the permissions or not. Thus we are unable to determine whether it is a benign skill or a masked skill just for collecting users' information. For instance, \textit{Liquor Emporium} is a coupon skill which requests the permissions of name, address, email and phone number. However, we invoked it multiple times during different time periods and it always failed to provide coupons. 10 out of the 11 potentially over-privileged skills are from the Shopping category and the other one is from the Food category.

(4) Amazon judges skills' legitimacy based on whether they respond differently with and without the requested permissions. However, whether the skills have truly used the information and how to use it to implement the functions is out of its scope. Among the legitimate skills, 5 skills' functions are found having no relations with requested permissions. Skills normally use email and name to identify a user and provide customized welcome information. However, none of these skills have used the information in this way. They are highly potentially  designed just for collecting users' information, e.g., \textit{Daddy Saturday} gives ideas on what to do with daddy on weekends. It requests sensitive permissions of address, email and full name. However, only the address is used to get the weather and the skill gives advice according to the weather. Users have no idea what the skill has done with their emails and full names. Similarly, \textit{Biggest Fan} is a quiz game that asks for name, email and phone number, but none of the information is used in its responses. The legitimate skills with over-used permissions are shown in Table~\ref{over-used}.

\begin{table}[htbp]
	\setlength{\abovecaptionskip}{0cm}
		\setlength{\belowcaptionskip}{-0.2cm}

	\caption{Legitimate skills with over-used permissions}
	\begin{center}
		\scalebox{0.9}{
		\begin{tabular}{p{71pt}ccc}
			\toprule  
			Skill name&Category&Requested&Over-used\\
			&&sensitive&sensitive\\
			&&permissions&permissions\\
			
			\midrule
			Daddy saturday&Life style&1, 2, 3&1, 3\\
			Aawaz biggest fan&Game&1, 3, 4&1, 3, 4\\
			Aawaz fan club&Game&1, 3, 4&1, 3, 4\\
			Susu state quiz&Game&1, 2, 3&1, 3\\
			KIG&Game&1, 3&1, 3\\
			\bottomrule 
			\multicolumn{4}{l}{	1: Name 2: Address 3: Email 4: Phone number}
			
		\end{tabular}
		}
		\label{over-used}
	\end{center}
\end{table}

\vspace{-0.2cm}
\subsection{Same invocation names} 

Zhang \textit{et al.} \cite{wangxiaofeng} reported that a malicious skill may hijack a benign skill by registering a similar invocation name. For instance, when users say ``Alexa, open \textit{Capital One} please'', which normally opens the skill \textit{Capital One}, may trigger a malicious skill \textit{Capital One Please}. When multiple skills' invocation names are similar, the skill whose name has the longest string matching with the uttered name is invoked. However, what if skills have the exactly the same invocation names? Developers are allowed to use duplicate invocation names when creating a skill. The strategy to select the skill is still unknown when multiple skills share the same invocation name.

\begin{figure}[htbp]

\setlength{\abovecaptionskip}{0.1cm}
		\setlength{\belowcaptionskip}{-0.1cm}

	\centering
	\includegraphics[width=.45\textwidth]{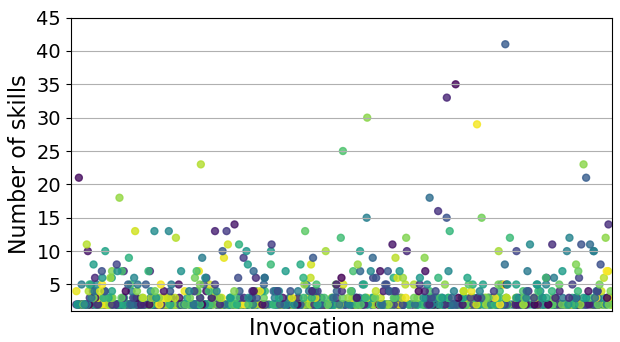}
	\caption{The same invocation name shared by skills. The top blue spot represents the invocation name is shared by 41 skills.}\label{same_invocation_name_colorful}
\end{figure}

In our dataset, 1,260 invocation names are used by more than one skill. The most popular one is \textit{space facts}, which is shared by 41 skills. These skills tell users facts of the planet or universe. The second most popular name shared by 35 skills is \textit{whose turn}. The skills are capable to help users decide whose turn to perform a task, e.g., taking out the trash or paying the bill. We also make a statistic of the duplication frequency. 822 invocation names are shared by 2 skills, which account for 65\% of all duplicated names. 48 invocation names are shared by over 10 skills, indicating that the duplication of invocation names is truly common. The detail is shown in Fig.~\ref{same_invocation_name_colorful}.

To find out Alexa's strategy for picking skills that have the same invocation names, we performed multiple sets of controlled variable experiments. We enabled several skills that have the same invocation name and say the invocation name 20 times to count the number of times each skill is invoked. Finally, we have the following observations: (1) If users have not enabled any skills that have this invocation name ever, Alexa provides the most popular one in the market; (2) If only one skill is enabled, then it will be invoked; (3) If more than one skill is enabled, on condition that the skills have the same rating, the earliest and the latest enabled ones tend to have higher priority. Besides the order of enabling and rating, popularity of developers and invocation history may also be taken into consideration, although we were unable to verify the latter factors for the lack of access to the data (e.g., the number of users of each skill). 

\vspace{-0.2cm}
\subsection{Developers} 
Next, we make a survey study of skill developers. We find that 3,763 developers have published more than one skill. The top 3 developers are \textit{InfoByVoice} with 1,625 skills, \textit{Patch.com} with 902 skills, and \textit{Rhall} with 624 skills. \textit{InfoByVoice} provides information about different churches. \textit{Rhall}'s skills are encyclopedia that can tell users the facts about fruit, geography, animals, etc. \textit{Patch.com} provides local news and events. Among the 3,763 developers, 93\% of them have less than 10 skills. It is challenging to detect malicious skills from a large number of skills. Therefore, if Amazon has found a malicious skill, it is highly recommended to check the developer's other skills because the developer may have made more. For example, the over-privileged skills mentioned in Section~\ref{over-per}, \textit{Aawaz fan club} and \textit{Aawaz biggest fan}, are both developed by \textit{Aawaz}. 1 over-privileged skill as well as 9 potentially over-privileged coupon skills were developed by \textit{Blutag}, which is an advertisement company that enables retail companies to recommend their brands by voice assistant. The number of developers with different numbers of skills are shown in Table~\ref{developer}.

\begin{table}[htbp]
	\setlength{\abovecaptionskip}{0cm}
		\setlength{\belowcaptionskip}{-0.2cm}

	\caption{Number of skills published by the same developer}
	\begin{center}
	\scalebox{0.9}{
		\begin{tabular}{cc}
			\toprule  
			Number of skills& Number of developers\\
			\midrule
			2 - 9&3548\\
			10 - 49&183\\
			50 - 99&16\\
			100 - 499&12\\
			500 - 999&2\\
			$\geq$ 1000&1\\
			
			\bottomrule 
			
		\end{tabular}
		}
		\label{developer}
	\end{center}
\end{table}

\vspace{-0.4cm}
\subsection{Google actions and Baidu skills} 
So far our work has focused on exploring the vulnerabilities against Alexa. A natural question arises: \textit{are these vulnerabilities unique to Alexa or also common to other voice assistants?} If other voice assistants adopt the same development model as Alexa, do they have to face the same problems? To answer these questions, we further looked into Google assistant and Baidu assistant, which have the second largest market share of smart speakers in the world \cite{google-market-share} and the largest market share in China \cite{Baidu-market-share}, respectively. 
Skills are called \textit{actions} by Google.
It turns out that the development models of Google actions and Baidu skills are quite similar to that of Alexa skills. Their frontend and backend are also separated, so essentially they face the same dilemma as Alexa does.

In particular, for Google actions, developers can host the frontend in Dialogflow and deploy the backend code in Firebase. Similarly, Baidu skills' frontend is hosted in DuerOS console and the backend is deployed in Baidu cloud. We find that (1) both Google actions and Baidu skills are vulnerable to the hidden code-manipulation and hidden content-manipulation attacks. Similar to Alexa skills, the backend code and content from RSS links can be manipulated after the skills have passed the vetting; (2) Both Google actions and Baidu skills suffer the risk of over-privileged resource access. Google actions can only access two types of users' information by permissions: name and location. In Baidu skills, the permissions related to users' privacy include location and user profile, which contains public information such as nickname and portrait. In other words, the only private information of the users that can be obtained by Baidu skills' developer is users' locations. 
The less information a voice assistant accesses the less risky it is.

Google and Baidu, however, have a more relaxed vetting process, which makes the security situation remain serious. Firstly, 
a skill can easily request permissions even though the skill's function does not need them. The developer even does not have to make an excuse to request the permissions, as we did with Alexa skills for simulating attacks. We have developed a Google action and a Baidu skill that only tell jokes, but they requested the two permissions. They both passed the vetting and successfully went online. Then, we enabled these skills and our locations, names or nicknames information in the skills was sent to our server (in our short-period test, no one else used our skills). What's more, Google action does not have the permission checkbox to intuitively remind users and reviewers the requested permissions. The request is fully implemented by backend code. Due to the code-manipulation attack, the attacker can manipulate the requested permissions after the action is published. With user's name and location as well as sensitive question ``Are you home alone'', attackers are still able to carry out attacks in the real world. A comparison over the security of skills from these three voice assistants is shown in Table~\ref{comparison-of-assistants}.

\begin{table}[htbp]
	\setlength{\abovecaptionskip}{0cm}
		\setlength{\belowcaptionskip}{-0.2cm}

	\caption{Comparison of three voice assistants}
	\begin{center}
	\scalebox{0.9}{
		\begin{tabular}{cp{80pt}ccc}
			\toprule  
			&Item&Amazon&Google&Baidu\\
			&&Alexa&Assistant&Assistant\\
			\midrule
			1&Vulnerable to Attack 1&Yes&Yes&Yes\\
			2&Vulnerable to Attack 2&Yes&Yes&Yes\\
			3&Vulnerable to Attack 3&Yes&Yes&Yes\\
			4&Permission checkbox&Yes&No&Yes\\
			5&Number of sensitive permissions&4&2&1\\
			6&Vetting&Strict&Loose&Medium\\
			
			\bottomrule 
			\multicolumn{5}{l}{Attack 1: Over-privileged resource access}\\
			\multicolumn{5}{l}{Attack 2: Hidden code-manipulation}\\
			\multicolumn{5}{l}{Attack 3: Hidden content-manipulation}\\
			\multicolumn{5}{l}{*The value ``Strict'', ``Medium'' and ``Loose'' are}\\
			\multicolumn{5}{l}{relative among the 3 assistants}\\
			
		\end{tabular}
		}
		\label{comparison-of-assistants}
	\end{center}
\end{table}

\vspace{-0.3cm}
\section{Discussions and Countermeasures}  

\subsection{Discussions} 
Above we have conducted a thorough survey study of Amazon skills market and confirmed that over-privileged skills did exist despite the strict vetting process. Although we have successfully implemented the hidden-code manipulation attack and hidden-content manipulation attack and such attacks were not detected for a long time, we did not find other malicious skills in the market exploiting these two vulnerabilities. 
We further tested the skills with low user rating (score under 2.0) to see whether the low scores were due to their malicious behaviors. We found that users' main complaints included (1) the skill is unavailable; (2) the contents do not update in time; and (3) it keeps users waiting for a long time. However, due to the hidden-code manipulation and hidden-content manipulation attacks, the function and content of a skill may frequently change. Attackers' temptation of stealing users' private information or spreading threatening messages can be easily covered up if the malicious functions only last for a period of time. 

The fact that we did not find malicious skills utilizing hidden-code manipulation and hidden-content manipulation vulnerabilities does not mean that they do not exist. We have clearly shown the feasibility and severity of the attacks. Effective measures must be taken to prevent such attacks in the first place and further detect malicious skills in the market.    

\vspace{-0.2cm}
\subsection{Deal with over-privileged resource access} 
Amazon only tests whether a skill responds differently with and without permissions.  For a skill elaborately designed to collect user's personal information with a reasonable excuse, e.g., our \textit{susu assistant} skill, it is difficult to recognize its real purpose. Even if Amazon could access the source code and code analysis were feasible, the behavior of transmitting personal information to a remote server cannot be simply treated as malicious. It depends on the way the information is used by the functions of the skill. In some cases, the personal information \textit{need not} be stored on the server side (e.g., in the developer's database); instead, it can be directly used in the services right after the skill gets it. For instance, once a skill gets the name of the user, e.g., \textit{Mike}, it can immediately use the name in the customized welcome message ``Hello Mike, welcome to the skill.'' There is also no need to store users' information on the server side if skills only provide local news or weather information. However, in some other cases, storing users' information on the server side for providing services becomes necessary. In game or story skills, name and email are usually used to identify a user. The information is required to be stored in the developers' database so that whenever users come back they are able to resume the game or story. Thus, the behavior of ``transmitting users' information to remote servers'' is too complicated to be simply judged as benign or malicious, let alone Amazon cannot access the source code. 

Currently, the responsibility is handed over to users whether to give the permissions or not. However, many users may simply enable the skill without paying attention to the default-checked permission checkbox. It is necessary to set the default status of the checkbox unchecked and encourage users to check by themselves. Although it may affect user experience, it will slow down the enabling process and earn users more time to decide. 

Finally, the purpose of requesting each permission should be described in the skill's description instead of shown in the privacy policy. Currently, most of the skill descriptions are very short without mentioning permissions. As shown in Table~\ref{description_length} and Figure~\ref{average_length}, 54.4\% skills' descriptions contain less than 50 words. The average length of descriptions in News and Weather category is only 41. Among the 16 skills that have requested four sensitive permissions, only 6 mentioned the purposes of permissions in the descriptions. For skills that requested three sensitive permissions, the ratio is only 4 out of 23. 
\begin{table}[htbp]

	\setlength{\abovecaptionskip}{0cm}
		\setlength{\belowcaptionskip}{-0.2cm}

	\caption{Distribution of descriptions' length}
	\begin{center}
	\scalebox{0.9}{
		\begin{tabular}{cc}
			\toprule  
			Descriptions' length&Percentage of skills\\
            (Number of words)&\\			
			\midrule
			\textless 50&54.4\%\\ 
            50 - 99&26.8\%\\
            100 - 149&7.8\%\\
            150 - 199&4.1\%\\
            $\geq$ 200&6.9\%\\

			\bottomrule 
			
		\end{tabular}
		}
		\label{description_length}
	\end{center}
\end{table}

\begin{figure}[htbp]
\setlength{\abovecaptionskip}{0.1cm}
		\setlength{\belowcaptionskip}{-0.2cm}

	\centering
	\includegraphics[width=.45\textwidth]{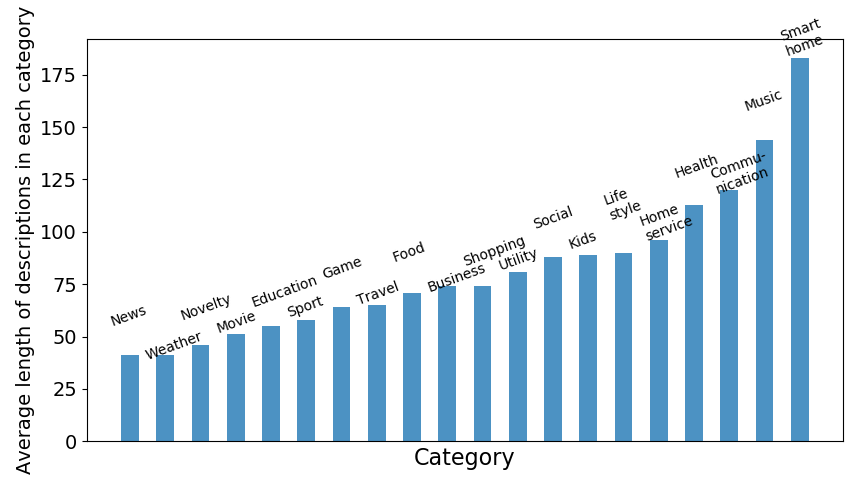}
	\caption{Average length of descriptions in each category}\label{average_length}
\end{figure}

\vspace{-0.2cm}
\subsection{Deal with hidden code-manipulation} 
In the current Alexa ecosystem, the backend code is not accessible to Amazon when it is hosted on the developer's own server. 
One reason could be to avoid code leakage and copyright infringement problems if all source code is open to Amazon.
However, to prevent hidden code-manipulation, backend code checking should be imperative whenever a skill provides an updated version.  Moreover, a developer should list \textit{all} the questions that may be asked by the skill in its description or in the frontend code. Currently, most skills' descriptions are very brief with few questions listed. Therefore, users may enable a skill without clearly understanding its exact functions and the utterances. They hold interests in exploring the functions in the skill by real-time conversations. However, if the conversations are changed to ``Are you home alone'' with an appropriate context, users may let the guard down and offer the real answers. In contrast, if users or Amazon are aware of all the questions of a skill, it will be obviously abnormal when the attacker changes ``Do you want to hear a joke'' to a completely different question.

Next, we provide more details on how to extend the above idea to detect the additional sensitive questions by manipulating the backend code. Here we assume that developers have provided all the questions to be asked by their skills, and the list is known to Amazon.
We first define a sensitive question list as the blacklist, and then automatically identify sensitive questions in a skill by measuring how similar they are with those in the blacklist.  
When the skill changes some of the questions, Amazon will get a new question list. The only work required is to estimate whether the changed or newly added questions in the list are sensitive questions or common questions. 

\begin{figure}[htbp]

\setlength{\abovecaptionskip}{0.1cm}
		\setlength{\belowcaptionskip}{-0.2cm}

	\centering

	\includegraphics[width=.45\textwidth]{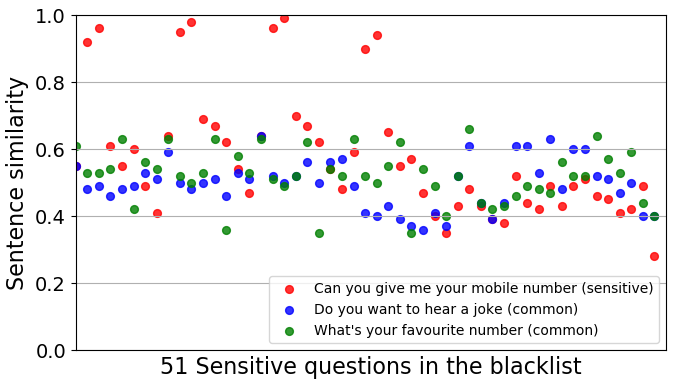}
	\caption{The similarities between three sample sentences and the blacklist using BERT model. The first sentence is a sensitive question that has a high similarity with the ones in the blacklist. Although the third sentence matches most of the words with the sensitive question ``What's your phone number'', we are still able to distinguish it from those in the blacklist.}\label{sentence_similarity}
\end{figure}

We come up with an initial blacklist that contains 51 expressions related to users' privacy, including their names, addresses, emails, phone numbers, genders, ages, family members and whether they are alone at the moment. These personal questions are considered to be of potential interest to attackers. Although it is probably incomplete, our mechanism can be easily tuned when adding new samples in the blacklist. Then we make use of a Natural Language Processing (NLP) tool called Sentence Transformers \cite{Sentence-Transformers}. It provides several popular pre-trained NLP models on the SNLI \cite{SNLI} and MultiNLI \cite{MultiNLI} datasets, including BERT \cite{BERT}, RoBERTa \cite{RoBERTa} and DistilBERT \cite{DistilBERT}. Sentence Transformers produces semantically meaningful sentence embeddings. It then computes the cosine similarity between the old/new question list and the blacklist, and determines whether any newly added questions are sensitive based on a similarity threshold. If so, the skill needs more attention and may need further investigation. 
Here the selection of similarity threshold is important to accurately distinguish sensitive questions and common questions.   
Based on our empirical study, the threshold value of 0.8 worked well. Since we have considered different expressions of personal questions and precisely defined the blacklist, only the ones have high similarity (above 0.8) are regarded as sensitive. The high threshold avoids false positives which happen when the majority of words in a common sentence and a sensitive sentence are the same. For example, the common question ``What's your favourite number'' and sensitive question ``What's your phone number'' have the similarity score 0.53; the sensitive question ``Can you give me your mobile number'' and the one in the blacklist `` Could you tell me your phone number'' have similarity score 0.96. 
The blacklist is shown in Appendix ~\ref{appendix_blacklist} and an example is shown in Fig.~\ref{sentence_similarity}.

\vspace{-0.2cm}
\subsection{Deal with hidden content-manipulation} 
To deal with hidden content-manipulation, Amazon is supposed to pay more attention to the unofficial news feeds, e.g., personal homepage, and increase the frequency of reviews. One obvious solution is to calculate the similarity between the description and content so as to verify whether the skill provides what it has promised. However, two challenges arise: (1) a large number of skill descriptions are as brief as one sentence, e.g., `` Five minutes of NPR news, updated hourly.'', which may not contain much useful information;  (2) Detecting fake news and rumors has become a special research field which is still facing a great challenge. It may be possible to tell when attackers manipulate the content from a specific area to another specific area, e.g., from recipes to pornographic messages.  However, there is still a long way to go to effectively and efficiently distinguish fake news from real news since news covers a wide range of subjects. Existing literature\cite{DBLP:conf/acl/Wang17,DBLP:conf/aaaiss/HinkelmannAC19,DBLP:journals/sigkdd/ShuSWTL17,DBLP:journals/corr/abs-1905-12616,DBLP:journals/corr/abs-1710-08528} and fact-checking website \cite{factchecking_website} provide some good ideas. We leave the solutions for our future work. 

\vspace{-0.2cm}
\section{Related Work}  

Most existing work on security issues of voice assistants focus on voice attacks, and some others are based on analyzing the encrypted traffic patterns to reveal the assistant's behaviors.

Bispham \textit{et al.} \cite{DBLP:conf/icissp/BisphamAG19}, Carlini \textit{et al.} \cite{DBLP:conf/uss/CarliniMVZSSWZ16} and Zhang \textit{et al.} \cite{DBLP:conf/ccs/ZhangYJZZX17} showed that Alexa can be attacked by nonsensical word sequences. Attackers may hide commands in nonsensical sounds, of which the meaning is opaque to humans, to take control of Alexa. To deal with the limitation that most voice attacks require a speaker close to the target to broadcast the obfuscated sound (e.g., DolphinAttack \cite{DolphinAttack}), Yuan \textit{et al.} \cite{DBLP:conf/globecom/YuanCWCZHM18} used signal broadcasting to compromise radio or TV and further implemented a remote voice control attack to control Alexa. Abdullah \textit{et al.} \cite{DBLP:conf/ndss/AbdullahGPTBW19} showed that the existed hidden voice attack targeting on voice processing systems highly depends on white-box knowledge of a specific system model or hardware, e.g., microphones and speakers. They broke the dependencies by exploiting the fact that multiple source audio samples have similar feature vectors when transformed by acoustic feature extraction algorithms. They created four classes of perturbations that can make unintelligible audio and successfully attack against targets. Lei \textit{et al.} \cite{DBLP:conf/cns/LeiTLLX18} found that Alexa devices took voice commands even if no people were around. Remote attackers can send commands and home burglary may take place. The authors believe that Alexa should add a physical presence based access control. They leveraged WiFi to detect indoor human motions to make sure Alexa can be invoked only when someone is at home. Similarly, Meng \textit{et al.} \cite{DBLP:conf/mobihoc/MengWZWZLL18} developed WiVo, which is a voice liveness detection system to distinguish the authentic voice command from a spoofed one via its corresponding mouth motions. Zhang \textit{et al.} \cite{wangxiaofeng} and Kumar \textit{et al.} \cite{DBLP:conf/uss/KumarPMHMBB18} both discovered the voice squatting attack. If multiple skills have the same or similarly pronounced invocation names, Alexa will randomly pick one of them when the invocation name is called. Thus a malicious skill has a high probability of being invoked. Zhang \textit{et al.} \cite{ZhangXMYCG19}analyzed and evaluated the security of the succeeding component after voice assistant's Automatic Speech Recognition. They developed LipFuzzer to discover potential misinterpretation-prone spoken errors. They verified LipFuzzer's effectiveness on Amazon Alexa and Google Assistant.

There also exist related work on analyzing
the encrypted traffic. Jackson \textit{et al.} \cite{DBLP:conf/icccn/JacksonC18} extracted the encrypted TCP traffic between an Alexa device and Amazons servers. They applied machine learning approaches to classify the traffic into six categories: information, quotes, weather, directions, music, and unintelligible. Ford \textit{et al.} \cite{DBLP:journals/puc/FordP19} confirmed that the on/off button does prohibit audio recording and streaming to Alexa Voice Service by analyzing network traffic of two Echo Dots over a 21-day period. The traffic has a sharp increase only when the switch is on and users say the wake word ``Alexa''. Chung \textit{et al.} \cite{DBLP:journals/corr/ChungPL17} gave a detailed introduction of Alexa's ecosystem. They developed a new approach that combined cloud-native forensics with client-side forensics to support practical digital investigations.

In addition to the security research on voice assistants, there are a rich literature on IoT security. 
For example, Alrawi \textit{et al.} \cite{DBLP:conf/sp/AlrawiLAM19} evaluated the security posture of 45 home-based IOT devices from four aspects: the IoT device, the companion mobile application, the cloud endpoints, and the associated communication channels. They found much of the same issues discussed in the literature exist in IoT systems today. 
Montoya \textit{et al.} \cite{Montoya} proposed SWARD to deal with Denial-of-Sleep attack, where attackers continuously send wake-up tokens to deplete the battery of the IoT nodes. They proposed a protocol to mitigate these attacks that includes a novel solution to generate hard-to-guess wake-up tokens at every wake-up.
Montoya \textit{et al.} \cite{Montoya} studied a type of denial-of-sleep attacks to deplete the battery of IoT nodes. 	Ding \textit{et al.} \cite{DBLP:conf/ccs/DingH18} proposed IoTMON to that discovers the possible physical interactions and generates all potential interaction chains across applications in the IoT environment. Apthorpe \textit{et al.} \cite{DBLP:journals/corr/ApthorpeRF17} evaluated 4 IoT smart home devices 
and found that their network traffic rates can reveal potentially sensitive user interactions even when the traffic is encrypted.

Celik \textit{et al.} \cite{DBLP:conf/usenix/CelikMT18} presented SOTERIA, a static analysis system for validating whether an IoT app or IoT environment adheres to identified safety, security, and functional properties. In their another work \cite{CelikTM19}, they noticed that the interactions among IOT devices are often the real cause of security violations. To protect users from unsafe device states, they presented a policy-based enforcement system to monitor the behavior of IoT and trigger-action platform apps.
Zhang \textit{et al.} \cite{DBLP:conf/ccs/ZhangMLZZZ18} proposed HoMonit by leveraging side-channel inference capabilities to monitor SmartApps from encrypted wireless traffic. They compared the SmartApps activities inferred from the encrypted traffic with their expected behaviors dictated in the source code or UI interfaces. 
They analyzed 181 official SmartApps and found 60 malicious SmartApps which either performed over-privileged accesses to smart devices or conducted event-spoofing attacks. 
Fernandes \textit{et al.} \cite{DBLP:conf/uss/FernandesPRSCP16} developed FlowFence to figure out how IoT apps use users' sensitive data by explicitly embedding data flows and the related control flows within the apps. Jia \textit{et al.} \cite{Jia} and Tian \textit{et al.} \cite{DBLP:conf/uss/TianZLWUGT17} applied context-based permission control to deal with the over-privileged IoT apps.

Independent from our work, Guo et al. \cite{guoskillexplorer} also focused on privacy issues of Alexa skills. They found that many skills request users' private information without following developer specifications. Our work, however, is the first systematic work that discovers three novel attacks and explains the root reasons of vulnerabilities as well as providing suggestions and solutions to fix the problems.

\vspace{-0.2cm}
\section{Conclusion} 
Our work presented three novel vulnerabilities and attacks on Amazon Alexa. Over-privileged resource access enables a skill to request more personal information than its function needs. Hidden code-manipulation attack manipulates backend code to bypass the vetting process and obtain sensitive information by real-time conversation. Hidden content-manipulation attack enables skills to inject harmful content, spread rumors and fake news which can mislead users and bring social panic. In addition, we gave detailed explanations of the root causes of vulnerabilities. We also made a thorough survey of 33,744 skills in Amazon market and found 57 over-privileged skills. Finally, to deal with the vulnerabilities in the popular voice assistant platforms, we proposed countermeasures and suggestions to better protect users' privacy.
	
	{

		\bibliographystyle{acm}
		\bibliography{ref}
		
	}

\section{Appendix}
\appendix
\section{Necessary elements for a custom skill} \label{appendix_elements}
The necessary elements for a custom skill are shown in Table ~\ref{elements}.

\begin{table}[htbp]
	\caption{Necessary elements for a custom skill}
	\begin{center}
	\scalebox{0.9}{
		\begin{tabular}{p{60pt}p{125pt}p{60pt}}
			\toprule  
			Element&Description&Example\\
			\midrule
		
            	Intents&Intents represent actions that users can do with the skill. Most skills support multiple functions and each function is represented by an Intent. The implementations of the intents are at the skill's backend.& \textit{JokeIntent} is defined to tell jokes.\\
			\midrule
			Utterances&Utterances are words and phrases users can say to invoke different intents. One Intent can have multiple utterances.& Developers can define ``joke'', ``tell me a joke'', ``I'd like to hear a joke'' as utterances of \textit{JokeIntent}. \\
			\midrule
			Invocation name&The invocation name is the identification of the skill. Users can speak to Alexa  ``Open \textit{Invocation name}'' to explicitly invoke the skill. It's worth noting that, Alexa allows duplicate invocation names, which means different skills can have the same invocation name. This may bring inconsistent user experience and the skill squatting attack as described in \cite{wangxiaofeng} and \cite{DBLP:conf/uss/KumarPMHMBB18}.&The invocation name of  \textit{Family jokes} is ``family jokes''.\\
			\midrule
            Permissions&The permissions that a skill requests to access users' personal information.&Name, email, address, phone number .etc.\\
            
            \midrule
            Endpoint link&A link to the skill's backend.&\\

			\bottomrule 
			
		\end{tabular}
		}
		\label{elements}
	\end{center}
\end{table}

\FloatBarrier

\section{The blacklist of sensitive questions} \label{appendix_blacklist}
The details of sensitive queations are shown in Table ~\ref{blacklist}.

\begin{table}[htbp]
	\caption{Sensitive question list}
	\begin{center}
	\scalebox{0.9}{
		\begin{tabular}{ll}
			\toprule  
			
			What's your&name,\\
			May I have your&phone/mobile number,\\
			Could you tell me your&address/location,\\
			I need your&email, age, gender\\
			\midrule
			\multirow{6}{*}{Are you}
			&home alone\\
			&with families/friends\\
			&married\\
			&over 18/an adult/a grown up\\
			&a boy/girl/teenager/kid\\
			\midrule
			\multirow{3}{*}{Do you have}
			&children/kids\\
			&boyfriend/girlfriend\\
			&husband/wife\\
			\midrule
			\multicolumn{2}{l}{Where are you}\\
			\multicolumn{2}{l}{How old are you}\\
			\bottomrule 
		\end{tabular}
		}
		\label{blacklist}
	\end{center}
\end{table}

\FloatBarrier
\section{Certification Requirements} \label{appendix_cer}
When a skill is submitted to the Alexa skill store, it must pass a certification process before it's published live to Amazon customers. The tests are shown in Table ~\ref{Certification}.

\begin{table}[htbp]
	\caption{Certification Requirements}
	\begin{center}
	\scalebox{0.9}{
		\begin{tabular}{p{240pt}}
			\toprule  
	    
			\textbf{Functional test} is to verify whether the skill can reflect the core functionality as described in the skill description. \\
			\midrule
		\textbf{Voice interface test} is to ensure that the skill offers polite and graceful prompts for good user experience. Any rude or impolite words should be prohibited.\\
				\midrule
			\textbf{Policy test} is to make sure the skill's content fits its targeted user group. For example, a child-oriented skill should not sell products or collect personal information, but these activities might be acceptable in a skill targeting adult users. The skill's content should not contain violence or unsettling messages. Advertising is not allowed in the skill except that the skill is specifically designed to promote a product or when the advertisements do not use Alexa's voice.  \\
				\midrule
\textbf{Security test} is a security requirement for skill's backend that should be a valid and trusted cloud-based service. It also contains several privacy requirements, e.g., ``The skill must not misuse customer personally identifiable information or sensitive personal information''\\

			\bottomrule 
			
		\end{tabular}
		}
		\label{Certification}
	\end{center}
\end{table}

\FloatBarrier

\section{The skills in our dataset} \label{appendix_dataset}
The details of skills in our dataset are shown in Table ~\ref{skills}.

\clearpage

\begin{table}[htbp]
	\caption{Description of skills in our dataset}
	\begin{center}
	\scalebox{0.9}{
		\begin{tabular}{lllll}
			\toprule  
			
			Category&Number&Max&Max&Max\\
			\quad&of&rating&rating&review\\
			&skills&number&&number\\
			\midrule
			Weather&750&284&5.0&108\\
			Communication&387&8,480&5.0&219\\
			Education&4,972&1,840&5.0&1,600\\
			Food&1,167&4,944&5.0&472\\
			Health&1,703&7,589&5.0&7,587\\
			Home service&215&123&5.0&75\\
			Kids&2,221&11,044&5.0&168\\
			Life style&5,241&20,444&5.0&7,587\\
			News&5,728&26,539&5.0&1,493\\
			Novelty&2,562&10,348&5.0&201\\
			Shopping&246&425&5.0&154\\
			Social&861&8,286&5.0&263\\
			Sport&1,285&1,269&5.0&65\\
			Movie&670&418&5.0&100\\
			Smart home&1,271&8,596&5.0&2,723\\
			Game&4,866&11,462&5.0&4,459\\
			Utility&211&1,816&5.0&73\\
			Music&548&17,343&5.0&17,334\\
			Business&1,354&198&5.0&117\\
			Travel&1,092&734&5.0&230\\
			Total&37,350&&&\\
			
			\bottomrule
			\multicolumn{5}{l}{The minimum rating number, rating and review number}\\
			\multicolumn{5}{l}{are all 0.}\\
			
		\end{tabular}
		}
		\label{skills}
	\end{center}
\end{table}

\end{document}